\documentclass[12pt]{article}
\usepackage{enumerate}
\usepackage{amsfonts}
\usepackage{amsmath}
\usepackage{amssymb}
\usepackage{amsthm}
\usepackage{latexsym}
\usepackage{color}

\def\H{\mathcal{H}}

\def\S{\mathfrak{S}}

\def\C{\mathfrak{C}}
\def\T{\mathfrak{T}}
\def\B{\mathfrak{B}}
\def\U{\mathfrak{U}}
\def\W{\mathfrak{W}}
\newcommand{\supp}{\mathrm{supp}}

\newcommand{\id}{\mathrm{Id}}
\newcommand{\Tr}{\mathrm{Tr}}

\newcommand{\shs}{\hspace{1pt}}
\newcounter{defin}  \newcounter{lemma}  \newcounter{theorem}
\newcounter{proposition} \newcounter{corol}  \newcounter{remark} \newcounter{example}
\newenvironment{lemma}{\par\refstepcounter{lemma}     \textbf{Lemma \thelemma.} }{\rm\par}
\newenvironment{theorem}{\par\refstepcounter{theorem}     \textbf{Theorem \thetheorem.}\ }{\rm\par}

\newenvironment{remark}{\par\refstepcounter{remark}     \textbf{Remark \theremark.}}{\rm\par}

\begin{document}

\title{Optimal form of the  Kretschmann-Schlingemann-Werner theorem\\ for energy-constrained quantum channels and operations}

\author{M.E.~Shirokov \\
Steklov Mathematical Institute, Moscow, Russia}
\date{}
\maketitle
%\vspace{50pt}
\begin{abstract}
It is proved that the energy-constrained Bures distance between arbitrary infinite-dimensional quantum channels is
equal to the operator \emph{E}-norm distance from \emph{any given} Stinespring isometry of one channel to the set of
all Stinespring isometries of another channel with the same environment.

The same result is shown to be valid for arbitrary quantum operations.
\end{abstract}

%\tableofcontents

\section{Introduction}

Quantum channels and operations play a central role in the theory of open quantum systems, quantum information theory, theory of quantum measurements and other  scientific directions related to quantum physics \cite{Kraus,H-SSQT,H-SCI,N&Ch,Wilde}. In the Schrodinger picture, quantum operation is a completely positive  trace-non-increasing linear map between Banach spaces of trace-class operators, quantum channel is a trace-preserving quantum operation.
The Stinespring theorem gives a useful characterization of quantum channels and operations \cite{St}. It implies  that any quantum operation (correspondingly, channel) $\Phi$ from a quantum system $A$ to a quantum system $B$ can be represented as
\begin{equation*}%\label{S-r}
\Phi(\rho)=\Tr_E V_{\Phi}\rho V^*_{\Phi},\quad \rho\in\T(\H_A),
\end{equation*}
where $V_{\Phi}$ is a contraction (correspondingly, isometry) from the input Hilbert space $\H_A$ into the tensor product of the output Hilbert  space $\H_B$ and some
Hilbert space $\H_E$ called \emph{environment} and $\T(\H_A)$ denotes the Banach spaces of trace-class operators on $\H_A$ \cite{H-SCI,Wilde}. The operator $V_{\Phi}$ is often called \emph{Stinespring operator} for $\Phi$.

In \cite{Kr&W+,Kr&W} Kretschmann, Schlingemann and Werner obtained  (by using some notions and arguments from \cite{B&Co}) continuity bounds for the map $V_{\Phi}\mapsto \Phi$ and the multivalued map $\Phi\mapsto V_{\Phi}$ w.r.t. the diamond-norm metric on the set of quantum operations and the operator-norm metric on the set of Stinespring operators.

In the case of finite-dimensional quantum channels these continuity bounds are obtained in \cite{Kr&W+} by proving that for any common Stinespring representation
\begin{equation}\label{c-S-r}
\Phi(\rho)=\Tr_E V_{\Phi}\rho V^*_{\Phi},\qquad \Psi(\rho)=\Tr_E V_{\Psi}\rho V^*_{\Psi}
\end{equation}
of any quantum channels $\Phi$ and $\Psi$ the following relation holds
\begin{equation}\label{KSW-1}
  \beta(\Phi,\Psi)=\min_{U\in \mathfrak{U}(\H_E)}\|V_{\Phi}-[I_B\otimes U]V_{\Psi}\|,
\end{equation}
where $\mathfrak{U}(\H_E)$ is the group of unitary operators on the space $\H_E$ and $\beta(\Phi,\Psi)$ is the Bures distance
between the channels $\Phi$ and $\Psi$ -- a metric on the set of channels equivalent to the diamond norm metric (see Section 2), $\beta(\Phi,\Psi)$ is defined as the maximal Bures distance between the states $\Phi\otimes\id_R(\rho)$ and $\Psi\otimes\id_R(\rho)$, $\H_R\cong\H_A$, $\rho$ is a state on $\H_{AR}$  (relation (\ref{KSW-1}) is obtained
in the proof of Theorem 1 in \cite{Kr&W+} in implicit form, it can be also obtained by using the arguments from the proof of Theorem 1 in \cite{B&Co}).
Since any Stinespring isometry of $\Psi$ with the same environment space $\H_E$
has the form  $[I_B\otimes U]V_{\Psi}$ for some $U\in\mathfrak{U}(\H_E)$ \cite[Ch.6]{H-SCI},
relation (\ref{KSW-1}) means that the Bures distance between any quantum channels $\Phi$ and $\Psi$ is equal to the
operator norm distance from any given Stinespring isometry  $V_{\Phi}$ of $\Phi$ to the set of all Stinespring isometries of $\Psi$ with the same environment.

This result is analogous to the famous Uhlmann theorem \cite{Uhlmann}, which can be formulated in terms of the Bures
distance between quantum states $\rho$ and $\sigma$ of a finite-dimensional quantum system $A$ as follows
$$
\beta(\rho,\sigma)=\min_{U\in \mathfrak{U}(\H_E)}\|\varphi-[I_A\otimes U]\psi\|,
$$
where $\varphi$ and $\psi$ are given purifications in $\H_{AE}\doteq\H_A\otimes\H_E$ of the states $\rho$ and $\sigma$ and
$\|\cdot\|$ denotes the norm in the Hilbert space $\H_{AE}$.

In the case of infinite-dimensional quantum channels and operations
it is shown in \cite{Kr&W} that
\begin{equation*}%\label{KSW-2}
  \beta(\Phi,\Psi)=\inf\|V_{\Phi}-V_{\Psi}\|,
\end{equation*}
where the infimum is over all common Stinespring representations
(\ref{c-S-r}). It is also shown that this infimum is attainable and that optimal Stinespring operators
$\widetilde{V}_{\Phi}$ and $\widetilde{V}_{\Psi}$ can be constructed as the operators from $\H_A$ into $\H_B\otimes(\H_E\oplus\H_E)=[\H_B\otimes\H_E]\oplus[\H_B\otimes\H_E]$
defined by settings
\begin{equation*}%\label{bar-v}
\widetilde{V}_{\Phi}|\varphi\rangle=V_{\Phi}|\varphi\rangle\oplus|0\rangle,\quad
\widetilde{V}_{\Psi}|\varphi\rangle=[I_B\otimes C]V_{\Psi}|\varphi\rangle\oplus \left[I_B\otimes\sqrt{I_{E}-C^*C}\right]\!V_{\Psi}|\varphi\rangle
\end{equation*}
for any $\varphi\in\H_A$,  where $C$ is a particular contraction in $\B(\H_E)$. So, to obtain a relation similar to (\ref{KSW-1}) it suffices to
show that this optimal contraction $C$ can be chosen in such a way that $\left[I_B\otimes\sqrt{I_{E}-C^*C}\right]\!V_{\Psi}=0$.
But it is not clear how to do this because of different subtleties appearing in the infinite-dimensional case (noncompactness of
the set of quantum states, nonexistence
of a saddle point of a minimax problem, nonexistence of a polar decomposition with a unitary operator for some trace class operators, etc.)
%This prevents to  construct optimal Stinespring operators within a given Stinespring representation.

Dealing with infinite-dimensional quantum channels and operations we have to replace the diamond norm distance between them
by some weaker measure of distinguishability which properly describes physical perturbations of such channels and operations \cite{SCT,W-EBN}.
One of such measures is induced by the energy-constrained diamond
norm of a Hermitian preserving linear map between  Banach spaces of trace-class operators introduced in \cite{SCT,W-EBN}  (this norm  differs from the eponymous norm used in \cite{Lupo,Pir}). Another, topologically equivalent measure is the energy-constained Bures distance $\beta^H_E(\Phi,\Psi)$
between quantum operations $\Phi$  and $\Psi$ defined as the maximal Bures distance between the operators $\Phi\otimes\id_R(\rho)$ and $\Psi\otimes\id_R(\rho)$, where $\rho$ runs over the set of states on $\H_{AR}$ with the mean energy $\Tr H\rho_A$ not exceeding some bound $E$ \cite{CID}.  If  the Hamiltonian $H$ of input system has  discrete spectrum of finite
multiplicity then the energy-constained Bures distance generates the strong convergence on the set  quantum channels and operations
\cite[Proposition 1]{CID}.
In fact, the role of $H$ can be played by any positive operator with this form of spectrum.\smallskip

The Kretschmann-Schlingemann-Werner theorem mentioned before (in what follows it will be called the KSW-theorem) is adapted in \cite{CSR,ECN}  for
the energy-constained Bures distance on the set of quantum operations. By modifying the arguments from \cite{Kr&W} it is shown that
\begin{equation*}%\label{KSW-rel}
\beta^H_E(\Phi,\Psi)=\inf\|V_{\Phi}-V_{\Psi}\|^H_E,
\end{equation*}
where the infimum is over all common Stinespring representations (\ref{c-S-r}) and $\|\cdot\|^H_E$ is
a special norm on the space $\B(\H_A,\H_{BE})$ of all linear bounded operators from $\H_A$ to $\H_{BE}$ generating the strong convergence topology on norm bounded subsets of $\B(\H_A,\H_{BE})$.  The value of $\|X\|^H_E$ is defined as the maximal value of $\|X\varphi\|$, where $\varphi$ runs over the set
of all unit vectors in $\H_A$ such that $\langle\varphi|H|\varphi\rangle\leq E$ (see details in Section 2).

The above assertion can be called the KSW theorem for energy-constrained quantum channels and operations.
The aim of this paper is to obtain the optimal form of this theorem by
proving a generalized version  of relation (\ref{KSW-1}) in this more general settings, namely, to show that for a \emph{given} common Stinespring representation (\ref{c-S-r}) of any quantum operations $\Phi$ and $\Psi$ the following relation holds
\begin{equation*}%\label{KSW+1}
  \beta^H_E(\Phi,\Psi)=\inf_{U\in \W_{\Psi}}\|V_{\Phi}-[I_A\otimes U]V_{\Psi}\|^H_E,
\end{equation*}
where $\W_{\Psi}$ is the set of all\emph{ partial isometries} in $\B(\H_E)$ such that $[I_B\otimes U^*U]V_{\Psi}=V_{\Psi}$.

This result is similar to the infinite-dimensional version of Uhlmann's theorem, which can be formulated in terms of the Bures
distance between operators $\rho$ and $\sigma$ in $\T_+(\H_A)$ as follows
\begin{equation}\label{U-th+}
\beta(\rho,\sigma)=\inf_{U\in \W_{\psi}}\|\varphi-[I_A\otimes U]\psi\|,
\end{equation}
where $\varphi$ and $\psi$ are given purifications in $\H_{AE}$ of the operators $\rho$ and $\sigma$ and
$\W_{\psi}$ is the set of all\emph{ partial isometries} in $\B(\H_E)$ such that $[I_B\otimes U^*U]|\psi\rangle=|\psi\rangle$ (we can not take
the infimum here over all unitary operators in $\B(\H_E)$, since in infinite dimensions
there exist partial isometries that can not be extended to unitary operators).

The first step in this direction was made in \cite[Theorem 2]{ECN}, where it is shown that
$$
\beta^H_E(\Phi,\Psi)\leq\inf_{U\in \W_{\Psi}}\|V_{\Phi}-[I_A\otimes U]V_{\Psi}\|^H_E\leq 2\beta^H_E(\Phi,\Psi)
$$
provided that the operator $H$ (used in definitions of $\beta^H_E$ and $\|\cdot\|^H_E$) has  discrete spectrum of finite
multiplicity. This observation is used essentially in the proof of the main result of this paper along with two-level approximation technique
which allows us to overcome the technical problems mentioned before.

\section{Preliminaries}

Let $\mathcal{H}$ be a separable  Hilbert space,
$\mathfrak{B}(\mathcal{H})$ the algebra of all bounded operators  on $\mathcal{H}$ with the operator norm $\|\cdot\|$ and $\mathfrak{T}( \mathcal{H})$ the
Banach space of all trace-class
operators on $\mathcal{H}$ with the trace norm $\|\!\cdot\!\|_1$. Let
$\mathfrak{S}(\mathcal{H})$ be  the set of quantum states (positive operators
in $\mathfrak{T}(\mathcal{H})$ with unit trace) \cite{H-SCI,Wilde}.

Denote by $I_{\H}$ the unit operator on a Hilbert space
$\mathcal{H}$ and by $\id_{\mathcal{\H}}$ the identity
transformation of the Banach space $\mathfrak{T}(\mathcal{H})$.

The fidelity of operators $\rho$ and $\sigma$ in $\T_+(\H)$ is defined as (see \cite[Appendix B]{F&R},\cite{Wilde})
\begin{equation}\label{fidelity}
 F(\rho,\sigma)=\|\sqrt{\rho}\sqrt{\sigma}\|^2_1=\left[\Tr\sqrt{\sqrt{\sigma}\rho\sqrt{\sigma}}\right]^2.
\end{equation}
Uhlmann's theorem \cite{Uhlmann} implies that
\begin{equation}\label{U-th}
 F(\rho,\sigma)=\sup_{\varphi,\psi}|\langle\varphi|\psi\rangle|^2=\sup_{\varphi}|\langle\varphi|\psi\rangle|^2,
\end{equation}
where the first supremum is over all purifications $\varphi$ and $\psi$ of the operators $\rho$ and $\sigma$, while the second one is over all purifications of the operator $\rho$
with a given  purification of $\sigma$ (it is assumed that the dimension of reference system is not less than $\dim\H$) \cite{H-SCI,Wilde}.

\smallskip

We will use the following observation.\smallskip

\begin{lemma}\label{F-l} \emph{Let $\varphi$ and $\psi$ be vectors in $\H_{AB}$, $\rho=\Tr_B|\varphi\rangle\langle\varphi|$ and
$\sigma=\Tr_B|\psi\rangle\langle\psi|$. Then
\begin{equation}\label{F-exp}
F(\rho,\sigma)=\max_{U\in\B_1(\H_B)}|\langle\psi|I_A\otimes U|\varphi\rangle|^2.
\end{equation}
where $\B_1(\H_B)$ is the unit ball in $\B(\H_B)$. If the state $\sigma$ is nondegenerate then any operator $\,U_0$ at which this
maximum is attained has the property
\begin{equation}\label{U-cond}
I_A\otimes U^*_0U_0|\varphi\rangle=|\varphi\rangle
\end{equation}
that means that the restriction of $\,U_0$  to the support of $\Tr_A|\varphi\rangle\langle\varphi|$ is an isometry.}
\end{lemma}
\smallskip

\emph{Proof.} The attainability of the maximum in (\ref{F-exp}) follows from the compactness of the unit ball $\B_1(\H_B)$ in the weak
operator topology \cite{B&R}.\smallskip

It follows from (\ref{U-th}) that the r.h.s. of (\ref{F-exp}) is not less than $F(\rho,\sigma)$.\smallskip

Let $|\phi\rangle=I_A\otimes U|\varphi\rangle$, $U\in\B_1(\H_B)$. Then
$\varrho=\Tr_B |\phi\rangle\langle\phi|=\Tr_B [I_A\otimes U^*U]|\varphi\rangle\langle\varphi|\leq\rho$.
So, the last formula in (\ref{fidelity}) and expression (\ref{U-th}) imply that
$$
|\langle\psi|I_A\otimes U|\varphi\rangle|^2=|\langle\psi|\phi\rangle|^2\leq F(\varrho,\sigma)\leq F(\rho,\sigma).
$$
This proves $"\geq"$ in (\ref{F-exp}) and shows that $F(\rho,\sigma)=F(\varrho,\sigma)$ provided that $U=U_0$.

If condition (\ref{U-cond}) is not valid then $\varrho\neq\rho$. Since $\varrho\leq\rho$, the assumed
nondegeneracy of $\sigma$ implies, by the last formula in (\ref{fidelity}),
that $F(\rho,\sigma)>F(\varrho,\sigma)$ contradicting to the previous observation. $\square$
\smallskip

The \emph{Bures distance} between operators  $\rho$ and $\sigma$ in $\T_+(\H)$ is defined as
\begin{equation}\label{B-d-s}
  \beta(\rho,\sigma)\doteq\inf_{\varphi,\psi}\|\varphi-\psi\|=\inf_{\varphi}\|\varphi-\psi\|=\sqrt{\|\rho\|_1+\|\sigma\|_1-2\sqrt{F(\rho,\sigma)}},
\end{equation}
where the first infimum is over all purifications $\varphi$ and $\psi$ of the operators $\rho$ and $\sigma$, while the second one is over all purifications of the operator $\rho$
with a given  purification of $\sigma$ (it is assumed that the dimension of reference system is not less than $\dim\H$) \cite{H-SCI,Wilde}.

The following relations  between the Bures distance and the  trace-norm distance hold (see \cite{H-SCI,Wilde}\cite[Appendix B]{F&R})
\begin{equation}\label{B-d-s-r}
\frac{\|\rho-\sigma\|_1}{\sqrt{\|\rho\|_1}+\sqrt{\|\sigma\|_1}}\leq\beta(\rho,\sigma)\leq\sqrt{\|\rho-\sigma\|_1}.
\end{equation}

A \emph{quantum  operation} $\,\Phi$ from a system $A$ to a system
$B$ is a completely positive trace non-increasing linear map from
$\mathfrak{T}(\mathcal{H}_A)$ into $\mathfrak{T}(\mathcal{H}_B)$. For any  quantum operation  $\,\Phi$ the Stinespring theorem (cf.\cite{St}) implies existence of a Hilbert space
$\mathcal{H}_E$ called \emph{environment} and  a contraction
$V_{\Phi}:\mathcal{H}_A\rightarrow\mathcal{H}_{BE}\doteq\mathcal{H}_B\otimes\mathcal{H}_E$ such
that
\begin{equation}\label{St-rep}
\Phi(\rho)=\mathrm{Tr}_{E}V_{\Phi}\rho V_{\Phi}^{*},\quad
\rho\in\mathfrak{T}(\mathcal{H}_A).
\end{equation}
If $\Phi$ is a trace preserving operation then it is called \emph{quantum channel}. In this case the contraction $V_{\Phi}$ in any Stinespring
representation (\ref{St-rep}) of $\Phi$ is an isometry \cite{H-SCI,Wilde}.\smallskip

The \emph{diamond norm} of a Hermitian preserving linear map $\Phi$ from $\T(\H_A)$ to $\T(\H_B)$ is defined as
\begin{equation*}%\label{d-norm}
\|\Phi\|_{\diamond}=\sup_{\rho\in\S(\H_{AR})}\|\Phi\otimes \id_R(\rho)\|_1,
\end{equation*}
where $R$ is a quantum system such that $\H_R\cong\H_A$ \cite{Kit,Wilde} (it is also called the norm of complete boundedness \cite{Paul,Wat}).  This norm induces the
metric on the set of quantum operations $\Phi$ and $\Psi$ from $A$ to $B$ called diamond-norm metric
which is used essentially in the study of finite-dimensional quantum channels and operations.\smallskip

The \emph{Bures distance} between quantum operations $\Phi$ and $\Psi$ from $A$ to $B$ is defined as
\begin{equation*}%\label{b-dist+}
\beta(\Phi,\Psi)=\sup_{\rho\in\S(\H_{AR})} \beta(\Phi\otimes \id_R(\rho),\Psi\otimes \id_R(\rho)),
\end{equation*}
where $\beta(\cdot,\cdot)$ in the r.h.s. is the Bures distance between operators in $\T_+(\H_{BR})$ defined in (\ref{B-d-s}) and $R$ is a quantum system such that $\H_R\cong\H_A$ \cite{Kr&W, B&Co}. It is a metric on the set of all quantum operations that can be expressed (by the formula similar to (\ref{B-d-s})) via the \emph{operational fidelity}
\begin{equation*}%\label{op-fid}
F(\Phi,\Psi)=\inf_{\rho\in\S(\H_{AR})} F(\Phi\otimes \id_R(\rho),\Psi\otimes \id_R(\rho))
\end{equation*}
introduced in \cite{B&Co}. Here $F(\cdot,\cdot)$ in the r.h.s. is the fidelity of operators in $\T_+(\H_{BR})$ defined in (\ref{fidelity}) and $R$ is a system such that $\H_R\cong\H_A$.

The Bures distance and the diamond-norm metric are related by the inequalities
\begin{equation}\label{B-rel}
\frac{\|\Phi-\Psi\|_{\diamond}}{\sqrt{\|\Phi\|_{\diamond}}+\!\sqrt{\|\Psi\|_{\diamond}}}\leq\beta(\Phi,\Psi)\leq\sqrt{\|\Phi-\Psi\|_{\diamond}}
\end{equation}
showing their equivalence on the set of all quantum operations \cite{Kr&W}. But the Bures distance is more convenient in some cases. It allows, in particular, to obtain tight and close-to-tight continuity bounds for basic capacities of quantum channels depending on their input dimension \cite{CID}.
\smallskip

%\begin{equation}\label{star}
%\textstyle\frac{1}{2}\|\Phi-\Psi\|_{\diamond}\leq \beta(\Phi,\Psi)\leq\sqrt{\|\Phi-\Psi\|_{\diamond}}.
%end{equation}

Let $H$ be any positive (semidefinite) densely defined  operator on $\H_A$. We will treat $H$ as  Hamiltonian (energy observable) of a quantum system $A$, so that
$\Tr H\rho$ is the mean energy of a state $\rho$ in $\S(\H_A)$  (the value of $\Tr H\rho$ (finite or infinite) is defined as $\sup_n\Tr P_n H\rho$, where $P_n$ is the spectral projector of $H$ corresponding to the interval $[0,n]$). We will assume for simplicity that the Hamiltonian $H$ is grounded, i.e. $\inf_{\|\varphi\|=1}\langle\varphi|H|\varphi\rangle=0$.\smallskip

For given $E>0$ consider the norm on the space $\B(\H_A,\H_B)$ of
all bounded linear operators from $\H_A$ to $\H_B$ defined as
\begin{equation}\label{ec-on+}
 \|X\|^H_E\doteq \sup_{\varphi\in\H^1_A, \langle\varphi|H|\varphi\rangle\leq E}\|X\varphi\|,
\end{equation}
where $\H_A^1$ is the unit sphere in $\H_A$. This norm
can be also defined as
\begin{equation}\label{ec-on}
 \|X\|^H_E\doteq \sup_{\rho\in\S(\H_A),\Tr H\rho\leq E}\sqrt{\Tr X\rho X^*}.
\end{equation}
The coincidence of (\ref{ec-on+}) and  (\ref{ec-on}) is established in \cite{ECN} by using the main result in \cite{W-Sh} which allows us to
show that the supremum in (\ref{ec-on}) can be taken over pure states only.\smallskip

The norm $\|\cdot\|^H_E$  called \emph{operator E-norm} in \cite{ECN} was introduced in \cite{CSR} to obtain
energy-constrained version of the KSW theorem (described in the Introduction).\smallskip

Definition (\ref{ec-on+}) shows the sense of the operator E-norm $\|\cdot\|^H_E$ (as a constrained version of the operator norm $\|\cdot\|$) while definition
(\ref{ec-on}) is more convenient for studying its analytical properties. In particular, definition
(\ref{ec-on}) allows us to show that (see \cite{CSR},\cite[Remark 1]{ECN})
\begin{itemize}
  \item the  function $\,E\mapsto[\|X\|^H_E]^2$ is concave on $\,(0,+\infty)$ for each $X\in\B(\H_A,\H_B)$;
  \item the function $\,E\mapsto\|X\|^H_E\,$ is nondecreasing and tends to $\|X\|\,$ as $\,E\to+\infty$.
\end{itemize}
These properties imply that
\begin{equation}\label{E-n-eq}
\|X\|^H_{E_1}\leq \|X\|^H_{E_2}\leq \sqrt{E_2/E_1}\,\|X\|^H_{E_1}\quad\textrm{ for any } E_2>E_1>0
\end{equation}
and any $X\in\B(\H_A,\H_B)$. The inequalities in (\ref{E-n-eq}) show the equivalence of all the norms $\|\!\cdot\!\|^H_{E}$, $E>0$, on $\B(\H_A,\H_B)$.
\smallskip
If the operator $H$ has discrete spectrum of finite multiplicity then the  norm $\|\cdot\|^H_E$  generates the strong operator topology on norm bounded subsets of $\B(\H_A,\H_B)$ \cite{CSR}.\smallskip

The \emph{energy-constrained  diamond norm} of a Hermitian preserving linear map $\Phi$ from $\T(\H_A)$ to $\T(\H_B)$ induced by the operator $H$ is defined as
\begin{equation*}%\label{ec-d-norm}
\|\Phi\|^H_{\diamond,E}=\sup_{\rho\in\S(\H_{AR}),\Tr H\rho_A\leq E}\|\Phi\otimes \id_R(\rho)\|_1,
\end{equation*}
where $R$ is a quantum system such that $\H_R\cong\H_A$ \cite{SCT,W-EBN} (this norm differs from the eponymous norm used in \cite{Lupo,Pir}).
If the operator $H$ has discrete spectrum of finite multiplicity then the metric induced by the norm $\|\cdot\|^H_{\diamond,E}$ generates the strong convergence
on the set  of quantum channels and operations from $A$ to $B$ \cite[Proposition 3]{SCT}.\footnote{The strong convergence of a sequence $\{\Phi_n\}$ of quantum operations to a quantum operation $\Phi_0$  means that $\lim_{n\rightarrow\infty}\Phi_n(\rho)=\Phi_0(\rho)$  for all $\rho\in\S(\H_A)$ \cite{H-SCI,CSR}.}\smallskip

The \emph{energy-constrained Bures distance}
\begin{equation}\label{ec-b-dist}
\beta^H_E(\Phi,\Psi)=\sup_{\rho\in\S(\H_{AR}), \Tr H\rho_A\leq E} \beta(\Phi\otimes \id_R(\rho),\Psi\otimes \id_R(\rho)), \quad E>0,
\end{equation}
between quantum channels $\Phi$ and $\Psi$ from $A$ to $B$  induced by the operator $H$ (where $\,R\,$ is an infinite-dimensional quantum system) is introduced in \cite{CID} for quantitative continuity analysis of information characteristics of energy-constrained infinite-dimensional quantum channels. Properties of the energy-constrained Bures distance are presented in Proposition 1 in \cite{CID}. It is shown, in particular, that for given channels $\Phi$ and $\Psi$ the following properties hold
\begin{itemize}
  \item the function $E\mapsto [\beta^H_E(\Phi,\Psi)]^2$ is concave on $(0,+\infty)$;
  \item function $E\mapsto\beta^H_E(\Phi,\Psi)$ is nondecreasing and tends to $\beta(\Phi,\Psi)$ as $E\to+\infty$.
\end{itemize}
These properties imply that
\begin{equation}\label{beta-eq}
\beta^H_{E_1}(\Phi,\Psi)\leq \beta^H_{E_2}(\Phi,\Psi)\leq \sqrt{E_2/E_1}\,\beta^H_{E_1}(\Phi,\Psi)\quad\textrm{ for any } E_2>E_1>0.
\end{equation}
and any quantum channels $\Phi$ and $\Psi$. The inequalities in (\ref{beta-eq}) show the equivalence of all the distances $\beta^H_{E}$, $E>0$, on the set of all quantum channels from $A$ to $B$.
\smallskip
The calculations of $\beta^H_E(\Phi,\Psi)$ for real quantum channels can be found in \cite{Nair}.
\smallskip

The energy-constrained Bures distance can be defined by formula (\ref{ec-b-dist}) for arbitrary quantum operations $\Phi$ and $\Psi$. By using the arguments from \cite{CID} it is easy to show that it  possesses the above stated properties on the set of all quantum operations.
\smallskip

The energy-constrained Bures distance and the metric induced by the energy-constrained diamond norm are related by the inequalities
\begin{equation}\label{B-rel+}
\frac{\|\Phi-\Psi\|^H_{\diamond,E}}{\sqrt{\|\Phi\|^H_{\diamond,E}}+\!\sqrt{\|\Psi\|^H_{\diamond,E}}}\leq\beta^H_E(\Phi,\Psi)\leq\sqrt{\|\Phi-\Psi\|^H_{\diamond,E}}
\end{equation}
(the  energy-constrained versions of the inequalities in (\ref{B-rel}))  showing their equivalence on the set of all quantum operations \cite{CID,ECN}. If the operator $H$ has discrete spectrum of finite multiplicity then  the distance $\beta^H_E$ generates the strong convergence on the set  of quantum channels and operations from $A$ to $B$ \cite[Proposition 1]{CID}.\smallskip

\begin{remark}\label{ec-b-dist-r} The supremum in (\ref{ec-b-dist}) can be taken only over pure states $\rho\in\S(\H_{AR})$.
This follows from the freedom of choice of $R$, which implies possibility to purify any mixed state in $\S(\H_{AR})$ by extending system $R$. We have only to note that
the Bures distance between operators in $\T_{+}(\H_{XY})$ defined in (\ref{B-d-s}) does not increase under partial trace: $\beta(\rho,\sigma)\geq \beta(\rho_X,\sigma_X)$ for any $\rho$ and $\sigma$ in $\T_{+}(\H_{XY})$ \cite{F&R,H-SCI,Wilde}. \smallskip
\end{remark}

We will use the following simple lemmas.\smallskip

\begin{lemma}\label{vsl} \emph{Let $\,\Phi$ be a quantum operation from $A$ to $B$ and
$\H_{\Phi}$ the minimal subspace of $\,\H_B$ containing the supports of all the operators $\,\Phi(\rho)$, $\rho\in\S(\H_A)$.
Then $\H_{\Phi}$ coincides with the support of $\,\Phi(\sigma)$ for any nondegenerate state $\sigma$ in $\S(\H_A)$.}\footnote{The support $\supp \varrho$
of a positive trace class operator $\varrho$ is the closed subspace spanned by its eigenvectors corresponding to nonzero eigenvalues.}
\end{lemma}\smallskip

\emph{Proof.} Let $\sigma=\sum_i \mu_i |i\rangle\langle i|$ be the spectral representation of $\sigma$ such that $\mu_{i+1}\leq\mu_i$ for all $i$. Suppose,
$\supp \Phi(\sigma)\neq \H_{\Phi}$. Then there is a state $\rho$ in $\S(\H_A)$ such that
$\supp \Phi(\rho)$ is not contained in $\supp \Phi(\sigma)$. Consider the sequence of states
$$
\rho_n=[P_n\rho]^{-1}P_n\rho P_n,\;\textrm{ where }\;P_n=\sum_{i=1}^n|i\rangle\langle i|.
$$
Since this sequence converges to the state $\rho$,  there exists $n_0$ such that
$\supp \Phi(\rho_{n_0})$ is not contained in $\supp \Phi(\sigma)$. On the other hand,
$\rho_{n_0}\leq P_{n_0}\leq \mu^{-1}_{n_0}\sigma$ and hence
$\Phi(\rho_{n_0})\leq \mu^{-1}_{n_0}\Phi(\sigma)$. This implies that $\supp \Phi(\rho_{n_0})\subseteq\supp \Phi(\sigma)$ contradicting to the choice of $n_0$. $\square$

\begin{lemma}\label{f-d-l} \emph{Let  $A$ be a linear operator on a finite-dimensional Hilbert space $\H$. Then
\begin{equation}\label{max-r}
\!\max_{U\in\B_1(\H)}|\Tr UA|=\Tr\sqrt{A^*A},
\end{equation}
where $\B_1(\H)$ is the unit ball in $\B(\H)$. If the operator $A$ is non-degenerate then  any operator $\,U_0$ at which this
maximum is attained is equal to $\,c[A^*A]^{-1/2}A^*$, $|c|=1$, that means that $\,U^*_0$ coincides (up to a scalar factor) with the unitary from the polar decomposition of $A$.}
\end{lemma}\smallskip

\emph{Proof.}  Equality (\ref{max-r}) obviously follows from the inequality $|\Tr UA|\leq\|U\|\|A\|_1$.\smallskip

Assume that the operator $A$ is non-degenerate and $U_0\in\B_1(\H)$ is such that $\Tr U_0A=\Tr\sqrt{A^*A}$.
Let $\{|i\rangle\}$ be a basis in $\H$ consisting of eigenvectors of $\sqrt{A^*A}$ and $\{\lambda_i\}$ the set of corresponding (nonzero)
eigenvalues. Denote by $W_A$ the operator $A[A^*A]^{-1/2}$. Then
$$
\sum_i\lambda_i\langle i| U_0W_A|i\rangle=\sum_i \langle i| U_0W_A\sqrt{A^*A}|i\rangle=\Tr U_0A=\Tr\sqrt{A^*A}=\sum_i\lambda_i.
$$
Since  $|\langle i| U_0W_A|i\rangle|\leq\|U_0W_A\|\leq1$, this implies that $\langle i|U_0W_A|i\rangle=1$ for all $i$. So, it follows from the condition
$\|U_0W_A\|\leq1$ that $U_0W_A=I_{\H}$. Hence, $U_0=W^*_A$. $\Box$

\section{The main result}

Let $\Phi$ and $\Psi$ be arbitrary quantum operations from $A$ to $B$ having common Stinespring representation
\begin{equation}\label{c-S-r+}
\Phi(\rho)=\Tr_E V_{\Phi}\rho V^*_{\Phi},\qquad \Psi(\rho)=\Tr_E V_{\Psi}\rho V^*_{\Psi},\qquad \rho\in\T(\H_A).
\end{equation}
Such common representation can be obtained from the individual Stinespring representations for $\Phi$ and $\Psi$
by using isometrical embedding of the environment space with smaller dimension into the another environment space.  \smallskip

Our main result is the following theorem containing a generalization of relation (\ref{KSW-1})
to the case of infinite-dimensional energy-constrained quantum channels and operations. \smallskip

\begin{theorem} \label{main}\emph{Let $H$ be an unbounded densely defined positive operator on a separable Hilbert space $\H_A$ having discrete spectrum $\{E_k\}_{k\geq0}$ of finite multiplicity, $E_0=0$, $\beta^H_E$ the energy-constrained Bures distance and $\,\|\cdot\|^H_E$ the operator E-norm induced by the operator $H$ (defined, respectively,  in (\ref{ec-b-dist}) and (\ref{ec-on+})).} \smallskip

\emph{Let $\,\Phi$ and $\,\Psi$ be quantum operations having common Stinespring representation (\ref{c-S-r+}). Then
\begin{equation}\label{main-rel}
\beta^H_E(\Phi,\Psi)=\inf_{U\in\mathfrak{W}_{\Psi}}\|V_{\Phi}-[I_B\otimes U] V_{\Psi}\|^H_E,
\end{equation}
where $\mathfrak{W}_{\Psi}$ is the set of all partial isometries $\,U\in\B(\H_E)$ such that $[I_B\otimes U^*U]V_{\Psi}=V_{\Psi}$.}

\smallskip

\emph{If $\,\dim\H_E<+\infty$ then the infimum in (\ref{main-rel}) is attainable and can be taken over the set of all unitary operators on $\H_E$.}
\end{theorem}\smallskip

\emph{Proof.} The proof below is technically difficult, although its basic idea is quite simple. It is described  in Remark \ref{main-r} after the proof, which also explains the need to use additional constructions and approximation steps for its strict implementation.\smallskip

Let $\sigma$ be a given nondegenerate state in $\S(\H_A)$ and $\{p_n\}$  a sequence in $(0,1)$ tending to zero as $n\to\infty$.
For each $\,n\in\mathbb{N}$ consider the function on the space $\B(\H_A,\H_{BE})$ of all bounded linear operators from $\H_A$ to $\H_{BE}$ defined as
\begin{equation}\label{e-c-n-n}
\|X\|_{E,n}^H=\sup_{\rho\in\C_{H,E}}\sqrt{\Tr X\Theta_n(\rho)X^*},\quad X\in\B(\H_A,\H_{BE}),
\end{equation}
where $\Theta_n(\rho)=(1-p_n)\rho+p_n\sigma$ and $\C_{H,E}$ is the set of states $\rho$ in $\S(\H_A)$ such that $\Tr H\rho\leq E$. Then
$\,[\|X\|_{E,n}^H]^2=(1-p_n)[\|X\|^H_E]^2+p_n\Tr X\sigma X^*\,$ and hence
\begin{equation}\label{E-n-ineq}
\left|[\|X\|_{E,n}^H]^2-[\|X\|_E^H]^2\right|\leq p_n\|X\|^2, \quad \forall X\in\B(\H_A,\H_{BE}).
\end{equation}

For each $\,n\in\mathbb{N}$ consider the quantity
\begin{equation}\label{beta-n}
\beta_{E,n}^H(\Phi,\Psi)=\sup_{\omega\in\S_n} \beta(\Phi\otimes \id_R(\omega),\Psi\otimes \id_R(\omega)),
\end{equation}
where $R$ is an infinite-dimensional quantum system, $\beta(\cdot,\cdot)$ in the r.h.s. is the Bures distance between operators in $\T_+(\H_{BR})$ defined in (\ref{B-d-s}),
and $\S_n$ is the set of all pure states $\omega$ in $\H_{AR}$ such that $\omega_A=\Theta_n(\rho)$ for some state $\rho$ in $\C_{H,E}$.
By using the well known relation between different purifications of a given state (and the invariance of the Bures distance under isometrical transformation
of both arguments) it is easy to show that
the supremum in (\ref{beta-n}) can be taken over the set $\{\omega_{\Theta_n(\rho)}\,|\,\rho\in\C_{H,E}\}$, where
$\omega_{\Theta_n(\rho)}$ is a given arbitrarily chosen purification of $\Theta_n(\rho)$ in $\S(\H_{AR})$.

Since the Bures distance is a metric on $\T_+(\H_{BR})$, for arbitrary states $\omega$ and $\tilde{\omega}$
in $\S(\H_{AR})$ we have
\begin{equation}\label{b-est}
\begin{array}{c}
\beta(\Phi\otimes \id_R(\tilde{\omega}),\Psi\otimes \id_R(\tilde{\omega}))\leq\beta(\Phi\otimes \id_R(\omega),\Phi\otimes \id_R(\tilde{\omega}))\quad\\\\
+\beta(\Phi\otimes \id_R(\omega),\Psi\otimes \id_R(\omega))+\beta(\Psi\otimes \id_R(\omega),\Psi\otimes \id_R(\tilde{\omega}))\\\\
\leq\beta(\Phi\otimes \id_R(\omega),\Psi\otimes \id_R(\omega))+2\sqrt{\|\omega-\tilde{\omega}\|_1},
\end{array}
\end{equation}
where the last inequality follows from the right inequality in (\ref{B-d-s-r}) and monotonicity of the trace norm
under action of a quantum operation.

Since $\|\rho-\Theta_n(\rho)\|_1\leq 2p_n$ for any state $\rho$ in $\C_{H,E}$, for any purification
$\omega_{\rho}$ of a state $\rho$ there is a purification $\omega_{\Theta_n(\rho)}$ of
the state $\Theta_n(\rho)$ such that $\|\omega_{\rho}-\omega_{\Theta_n(\rho)}\|_1\leq 2\sqrt{(2-p_n)p_n}\leq2\sqrt{2p_n}$
and vice versa \cite{H-SCI,Wilde}. Hence, by Remark \ref{ec-b-dist-r}, estimate (\ref{b-est}) implies that
\begin{equation}\label{b-n-ineq}
\left|\beta_{E,n}^H(\Phi,\Psi)-\beta^H_E(\Phi,\Psi)\right|\leq 2\sqrt[4]{8p_n},\quad \forall n.
\end{equation}

Assume first that $\dim\H_E=+\infty$ and that
\begin{equation}\label{b-as}
  \ker\Phi\otimes\id_R(\omega)=\{0\}\;\textrm{ for any state }\;\omega\in\S(\H_{AR}) \textrm{ such that } \ker\omega_R=\{0\}.
\end{equation}

It follows from (\ref{e-c-n-n}) that
\begin{equation}\label{w-inf}
\inf_{U\in\mathfrak{W}_{\Psi}}\|V_{\Phi}-[I_B\otimes U] V_{\Psi}\|_{E,n}^H=\inf_{U\in\mathfrak{W}_{\Psi}}\sup_{\rho\in\C_{H,E}}\sqrt{f_n(\rho,U)},
\end{equation}
where $f_n(\rho,U)=\Tr\Phi(\Theta_n(\rho))+\Tr\Psi(\Theta_n(\rho))-2\Re\Tr [I_B\otimes U]V_{\Psi}\Theta_n(\rho)V^*_{\Phi}$.\smallskip

We will show that  for each $n$ the infimum in the r.h.s. of (\ref{w-inf}) can be taken over the unit ball $\B_1(\H_E)$ of $\B(\H_E)$.

The unit ball $\B_1(\H_E)$ is a convex subset of $\B(\H_E)$ compact in the weak operator topology \cite{B&R}, while $\C_{H,E}$ is a convex subset of
$\S(\H_A)$ compact in the trace norm topology by the Lemma in \cite{H-c-w-c}. So, since the functions $U\mapsto f_n(\rho,U)$ and $\rho\mapsto f_n(\rho,U)$\break
are affine and continuous on $\B_1(\H_E)$  and $\C_{H,E}$ correspondingly (in the above topologies), Ky Fan's minimax theorem \cite[Theorem 3.1]{Simons} implies that
\begin{equation}\label{minmax}
\min_{U\in\B_1(\H_E)}\max_{\rho\in\C_{H,\!E}}\sqrt{f_n(\rho,U)}
=\max_{\rho\in\C_{H,\!E}}\min_{U\in\B_1(\H_E)}\sqrt{f_n(\rho,U)}.
\end{equation}

Let $\widehat{\Psi}(\varrho)=\Tr_B V_{\Psi}\varrho V^*_{\Psi}$ be a quantum operation from $A$ to $E$ complementary to the operation $\Psi$ (cf.\cite{H-c-ch}) and
$P_{\widehat{\Psi}}$ the projector onto the
minimal subspace of $\H_E$ containing the supports of all the states $\widehat{\Psi}(\varrho)$, $\varrho\in\S(\H_A)$.

Let $U_0$ be an operator in $\B_1(\H_E)$ at which the minimum in the l.h.s. of (\ref{minmax}) is attained and
$\rho_0$ a state in $\C_{H,E}$ at which the maximum  in the r.h.s. of (\ref{minmax}) is attained. We may assume that
$U_0=U_0P_{\widehat{\Psi}}$, since it is easy to see that $f_n(\rho,U)=f_n(\rho,UP_{\widehat{\Psi}})$ for any $U$ and $\rho$. Then $(\rho_0,U_0)$
is a saddle point of the function $f_n$ \cite{Simons}, i.e.
\begin{equation}\label{s-p}
f_n(\rho,U_0)\leq f_n(\rho_0,U_0)\leq f_n(\rho_0,U)\quad \forall \rho\in\C_{H,E},U\in\B_1(\H_E).
\end{equation}

We will show that the last inequality in (\ref{s-p}) implies that $U_0\in\W_{\Psi}$.\footnote{This is a crucial point of the proof,
see Remark \ref{main-r} below.} This will imply coincidence of the r.h.s. of (\ref{w-inf})
and the l.h.s. of (\ref{minmax}).

For any $\rho$ in $\C_{H,E}$ we have
$$
\min_{U\in\B_1(\H_E)}\sqrt{f_n(\rho,U)}=\!\sqrt{\Tr\Phi(\Theta_n(\rho))+\Tr\Psi(\Theta_n(\rho))-2\!\max_{U\in\B_1(\H_E)}\!|\Tr [I_B\otimes U]V_{\Psi}\Theta_n(\rho)V^*_{\Phi}}|
$$
Let $R$ be an infinite-dimensional system and $\rho$ a state in $\C_{H,E}$. It is easy to see that
\begin{equation}\label{a-eq}
\!\max_{U\in\B_1(\H_E)}\left|\Tr [I_B\otimes U]V_{\Psi}\Theta_n(\rho)V^*_{\Phi}\right|=\!\!\max_{U\in\B_1(\H_E)}\left|\langle V_{\Phi}\otimes I_R\shs\varphi^{\shs\rho}_{n}|[I_{BR}\otimes U] V_{\Psi}\otimes I_R \shs\varphi^{\shs\rho}_{n}\rangle\right|,\!\!
\end{equation}
where $\varphi^{\shs\rho}_{n}$ is any purification of the state $\Theta_n(\rho)$, i.e. a vector in $\H_{AR}$ such that $\Tr_R|\varphi^{\shs\rho}_{n}\rangle\langle\varphi^{\shs\rho}_{n}|=\Theta_n(\rho)$.
Since the vectors
$V_{\Phi}\otimes I_R\shs|\varphi^{\shs\rho}_{n}\rangle$ and $V_{\Psi}\otimes I_R\shs|\varphi^{\shs\rho}_{n}\rangle$ in $\H_{BER}$ are purifications of the operators
$\Phi\otimes \id_R(|\varphi^{\shs\rho}_{n}\rangle\langle\varphi^{\shs\rho}_{n}|)$ and $\Psi\otimes \id_R(|\varphi^{\shs\rho}_{n}\rangle\langle\varphi^{\shs\rho}_{n}|)$ correspondingly, Lemma \ref{F-l} in Section 2
shows that the r.h.s. of (\ref{a-eq}) is equal to
$$
\sqrt{F(\Phi\otimes \id_R(|\varphi^{\shs\rho}_{n}\rangle\langle\varphi^{\shs\rho}_{n}|),\Psi\otimes \id_R(|\varphi^{\shs\rho}_{n}\rangle\langle\varphi^{\shs\rho}_{n}|))}
$$
and hence
\begin{equation}\label{b-n}
\min_{U\in\B_1(\H_E)}\sqrt{f_n(\rho,U)}=\beta(\Phi\otimes \id_R(|\varphi^{\shs\rho}_{n}\rangle\langle\varphi^{\shs\rho}_{n}|),\Psi\otimes \id_R(|\varphi^{\shs\rho}_{n}\rangle\langle\varphi^{\shs\rho}_{n}|)).
\end{equation}
Since the state $\Theta_n(\rho)$ is non-degenerate, we may take the above vector $\varphi^{\shs\rho}_{n}$ in such a way that $\ker \Tr_A|\varphi^{\shs\rho}_{n}\rangle\langle\varphi^{\shs\rho}_{n}|=\{0\}$. Hence  $\ker\Phi\otimes \id_R(|\varphi^{\shs\rho}_{n}\rangle\langle\varphi^{\shs\rho}_{n}|)=\{0\}$
by assumption (\ref{b-as}). Thus, the second claim of Lemma \ref{F-l} in Section 2 shows that
the maximum in (\ref{a-eq}) can be attained only at an operator $U_{\rho}$ such that
$$
[I_{BR}\otimes U_{\rho}^*U_{\rho}]V_{\Psi}\otimes I_R \shs|\varphi^{\shs\rho}_{n}\rangle=V_{\Psi}\otimes I_R \shs|\varphi^{\shs\rho}_{n}\rangle.
$$
This  condition means that $U_{\rho}^*U_{\rho}\widehat{\Psi}(\Theta_n(\rho))=\widehat{\Psi}(\Theta_n(\rho))$.
Since $\ker\Theta_n(\rho)=\{0\}$, Lemma \ref{vsl} in Section 2 implies that the operator $\bar{U}_{\rho}=U_{\rho}P_{\widehat{\Psi}}$ is a partial isometry
such that $\bar{U}_{\rho}^*\bar{U}_{\rho}=P_{\widehat{\Psi}}$. By this observation with $\rho=\rho_0$  the right inequality in (\ref{s-p}) implies that $U_0=U_0P_{\widehat{\Psi}}\in\W_{\Psi}$.\smallskip

Thus, the coincidence of the r.h.s. of (\ref{w-inf})
and the l.h.s. of (\ref{minmax}) is proved. So, by using (\ref{b-n}) we obtain
\begin{equation}\label{w-inf+}
\begin{array}{l}
\displaystyle\inf_{U\in\mathfrak{W}_{\Psi}}\|V_{\Phi}-[I_B\otimes U] V_{\Psi}\|_{E,n}^H\\\\\qquad\displaystyle=\max_{\rho\in\C_{H,E}}\beta(\Phi\otimes \id_R(|\varphi^{\shs\rho}_{n}\rangle\langle\varphi^{\shs\rho}_{n}|),\Psi\otimes \id_R(|\varphi^{\shs\rho}_{n}\rangle\langle\varphi^{\shs\rho}_{n}|))
=\beta_{E,n}^H(\Phi,\Psi)
\end{array}
\end{equation}
for each $n$, where $\varphi^{\shs\rho}_{n}$ is a particular purification of $\Theta_n(\rho)$ and the last equality follows from definition (\ref{beta-n}) and the remark after it.\smallskip

It follows from relation (\ref{b-n-ineq}) that $\beta_{E,n}^H(\Phi,\Psi)$ tends to $\beta^H_E(\Phi,\Psi)$ as $n\rightarrow+\infty$.
Since $\|V_{\Phi}-[I_B\otimes U] V_{\Psi}\|\leq2$ for any $U\in\W_{\Psi}$, relation (\ref{E-n-ineq}) shows that
\begin{equation}\label{imp-l-r}
\lim_{n\to+\infty}\inf_{U\in\mathfrak{W}_{\Psi}}\|V_{\Phi}-[I_B\otimes U] V_{\Psi}\|_{E,n}^H=\inf_{U\in\mathfrak{W}_{\Psi}}\|V_{\Phi}-[I_B\otimes U] V_{\Psi}\|^H_E
\end{equation}

These limit relations and (\ref{w-inf+}) imply (\ref{main-rel}) provided that the quantum operation $\Phi$
satisfies condition (\ref{b-as}).

Assume now that $\Phi$ is an arbitrary quantum operation. Consider the sequence of quantum operations
$$
\Phi_n(\rho)=(1-p_n)\Phi(\rho)+p_n[\Tr\rho]\sigma,
$$
from $A$ to $B$, where $\sigma$ is a given nondegenerate state in $\H_B$ and $\{p_n\}$ a sequence
of numbers in $(0,1)$ tending to zero as $n\to+\infty$.  It is easy to see that all the operations $\Phi_n$ satisfy
condition (\ref{b-as}).

The sequence $\{\Phi_n\}$ strongly converges to the operation $\Phi$ and hence $\beta^H_E(\Phi_n,\Phi)$ tends to zero as $n\to+\infty$ \cite{CID,CSR}.  Since the space $\H_E$ is infinite-dimensional, Theorem 2B in \cite{ECN} implies existence of a sequence of operators $V_{\Phi_n}:\H_A\rightarrow\H_{BE}$ such that
$$
\Phi_n(\rho)=\Tr_E V_{\Phi_n}\rho V_{\Phi_n}^*\quad \textrm{and} \quad \|V_{\Phi_n}-V_{\Phi}\|^H_E\leq 2\beta^H_E(\Phi_n,\Phi)+1/n\quad \forall n.
$$
By the above part of the proof we have
\begin{equation}\label{n-eq}
\beta^H_E(\Phi_n,\Psi)=\inf_{U\in\mathfrak{W}_{\Psi}}\|V_{\Phi_n}-[I_B\otimes U] V_{\Psi}\|^H_E.
\end{equation}
Since $\beta^H_E(\Phi_n,\Psi)$ tends to $\beta^H_E(\Phi,\Psi)$ as $\,n\to+\infty\,$ and
$$
\left|\|V_{\Phi_n}-[I_B\otimes U] V_{\Psi}\|^H_E-\|V_{\Phi}-[I_B\otimes U] V_{\Psi}\|^H_E\right|\leq\|V_{\Phi_n}-V_{\Phi}\|^H_E\leq 2\beta^H_E(\Phi_n,\Phi)+1/n
$$
for any $U\in\mathfrak{W}_{\Psi}$, by passing to the limit $n\to+\infty$ in (\ref{n-eq}) we obtain (\ref{main-rel}).\medskip

Consider  now the case $\dim\H_E<+\infty$. Assume first that $P_{\widehat{\Psi}}(\H_E)=\H_E$, i.e. that the union of supports of all the states $\widehat{\Psi}(\varrho)$, $\varrho\in\S(\H_A)$,
is dense in $\H_E$ (where $\widehat{\Psi}(\varrho)=\Tr_B V_{\Psi}\varrho V^*_{\Psi}$ is the operation complementary to the operation $\Psi$). \smallskip

For given $\varepsilon>0$ let
$\Phi_{\varepsilon}(\rho)=\Tr_EV^{\varepsilon}_{\Phi}\rho [V^{\varepsilon}_{\Phi}]^*$ be a completely positive linear map from $\T(\H_A)$ to $\T(\H_B)$, where
$V^{\varepsilon}_{\Phi}=V_{\Phi}+\varepsilon V_{\Psi}$.
By modifying the arguments presented after (\ref{b-as}) we want to show that relation
(\ref{w-inf+}) with $\Phi= \Phi_{\varepsilon}$ holds for all sufficiently small $\varepsilon$ and all $n$. The only modification
consists in using another way to prove that the last inequality in (\ref{s-p}) implies that $U_0\in\W_{\Psi}$  (instead of the paragraph after (\ref{b-n})
-- the only place where the assumption (\ref{b-as}) was used).\smallskip

Note that  the l.h.s. of (\ref{a-eq}) with $\rho=\rho_0$ can be written as
\begin{equation}\label{max-rel}
 \max_{U\in\B_1(\H_E)}|\Tr UA_\varepsilon|,
\end{equation}
where
$$
A_\varepsilon\doteq\Tr_B V^{\varepsilon}_{\Phi}\Theta_n(\rho_0)V_{\Psi}^*=\Tr_B V_{\Phi}\Theta_n(\rho_0)V_{\Psi}^*+\varepsilon\widehat{\Psi}(\Theta_n(\rho_0)).
$$
Since the state $\Theta_n(\rho_0)$ is non-degenerate, $\widehat{\Psi}(\Theta_n(\rho_0))$ is a non-degenerate operator on $\H_E$ by Lemma \ref{vsl} in Section 2.
Using this it is easy to show the existence of $a>0$ such that $A_\varepsilon$ is a non-degenerate operator on $\H_E$ for all $\varepsilon\in(0,a)$.
Thus, Lemma \ref{f-d-l} in Section 2 implies that any operator $U_0$ at which the maximum in (\ref{max-rel}) is attained
is unitary. So, it belongs to the set $\W_{\Psi}$, since the assumption $P_{\widehat{\Psi}}(\H_E)=\H_E$ implies that $\W_{\Psi}$
is the set of all unitary operators on $\H_E$.

The same arguments as before allow us to show that the validity of (\ref{w-inf+}) with $\Phi= \Phi_{\varepsilon}$ for all $\varepsilon\in(0,a)$ and all $n$
implies the validity of (\ref{main-rel}) with $\Phi=\Phi_{\varepsilon}$ for all $\varepsilon\in(0,a)$.
One should only to note that $\Phi_{\varepsilon}$ is not an operation in general (since $V^{\varepsilon}_{\Phi}$ may not be a contraction).
The only places,
where we have used that $V_{\Phi}$ and $V_{\Psi}$ are contractions, are
estimate (\ref{b-n-ineq}) obtained by using inequality (\ref{b-est})
and the inequality before (\ref{imp-l-r}). But, since $\|V^{\varepsilon}_{\Phi}\|\leq 2\,$ for any $\varepsilon\in(0,1)$,
we see  that estimate (\ref{b-n-ineq}) and the inequality before (\ref{imp-l-r}) hold in this case with factor $2$ replaced by $4$. This does not violate
in any way the validity of all arguments using there inequalities.

Since
$$
\inf_{U\in\mathfrak{W}_{\Psi}}\|V_{\Phi}-[I_B\otimes U] V_{\Psi}\|^H_E\leq \inf_{U\in\mathfrak{W}_{\Psi}}\|V^{\varepsilon}_{\Phi}-[I_B\otimes U] V_{\Psi}\|^H_E+\varepsilon\|V_{\Psi}\|
$$
by the triangle inequality and the obvious upper bound for the norm $\|\cdot\|^H_E$, we have
$$
\inf_{U\in\mathfrak{W}_{\Psi}}\|V_{\Phi}-[I_B\otimes U] V_{\Psi}\|^H_E\leq \lim_{\varepsilon\to0^+}\inf_{U\in\mathfrak{W}_{\Psi}}\|V^{\varepsilon}_{\Phi}-[I_B\otimes U] V_{\Psi}\|^H_E=\lim_{\varepsilon\to0^+}\beta^H_E(\Phi_{\varepsilon},\Psi)=\beta^H_E(\Phi,\Psi),
$$
where the last limit relation is due to the strong convergence of $\Phi_{\varepsilon}$ to $\Phi$ as $\varepsilon\to0^+$.
This inequality implies the validity of (\ref{main-rel}), since the converse inequality is obvious.

Since $\W_{\Psi}$ coincides with the set $\U(\H_E)$ of all unitary operators on $\H_E$ due to the assumption $P_{\widehat{\Psi}}(\H_E)=\H_E$,
equality (\ref{main-rel}) can be written in the symmetrical form
\begin{equation}\label{main-rel+}
\inf_{U\in\U(\H_E)}\|V_{\Phi}-[I_B\otimes U] V_{\Psi}\|^H_E=\inf_{U\in\U(\H_E)}\|V_{\Psi}-[I_B\otimes U] V_{\Phi}\|^H_E=\beta^H_E(\Phi,\Psi).
\end{equation}

Assume now that $\Phi$ and $\Psi$ are arbitrary operations having common Stinespring representation (\ref{c-S-r+})
with finite-dimensional environment space $\H_E$.

Suppose first that
$\dim P_{\widehat{\Psi}}(\H_E)\geq \dim P_{\widehat{\Phi}}(\H_E)$. Then there is a unitary operator $\widetilde{U}$ such that
$\widetilde{U} P_{\widehat{\Phi}}\widetilde{U}^*\leq P_{\widehat{\Psi}}$. Consider the common Stinespring representation
\begin{equation}\label{c-S-r++}
\Phi(\rho)=\Tr_E \widetilde{V}_{\Phi}\rho \widetilde{V}^*_{\Phi},\qquad \Psi(\rho)=\Tr_E V_{\Psi}\rho V^*_{\Psi},\qquad \rho\in\T(\H_A),
\end{equation}
where $\widetilde{V}_{\Phi}=[I_B\otimes \widetilde{U}] V_{\Phi}$. Since the ranges of $\widetilde{V}_{\Phi}$ and $V_{\Psi}$
belong to the subspace $\H_B\otimes\H_{\widetilde{E}}$, where $\H_{\widetilde{E}}=P_{\widehat{\Psi}}(\H_E)$, we may replace
the space $\H_E$ in (\ref{c-S-r++}) by its subspace $\H_{\widetilde{E}}$. Thus, the above part of the proof (under the assumption $P_{\widehat{\Psi}}(\H_E)=\H_E$)
shows that (\ref{main-rel+}) holds with $\H_E$ and $V_{\Phi}$ replaced by $\H_{\widetilde{E}}$ and $\widetilde{V}_{\Phi}$ correspondingly. But this
obviously guarantees  the validity of (\ref{main-rel+}) which implies (\ref{main-rel}).

If $\dim P_{\widehat{\Psi}}(\H_E)< \dim P_{\widehat{\Phi}}(\H_E)$ then we may prove (\ref{main-rel+}) by repeating the above arguments with a permutation of roles
$\Phi$ and $\Psi$. \smallskip

The claim about attainability of the supremum in (\ref{main-rel})  follows from continuity of the function $U\mapsto \|V_{\Phi}-[I_B\otimes U] V_{\Psi}\|^H_E$ on $\B(\H_E)$, since $\W_{\Psi}$ is a compact  subset of $\B(\H_E)$ in the case $\dim\H_E<+\infty$. $\square$
\smallskip

Since any Stinespring operator of a quantum operation $\Psi$ with the  same environment space $\H_E$
has the form  $[I_B\otimes U]V_{\Psi}$ for some $U\in\mathfrak{W}_{\Psi}$ \cite[Ch.6]{H-SCI},
relation (\ref{main-rel}) means that \emph{the energy-constrained Bures distance between quantum operations $\Phi$ and $\Psi$ is equal to the
operator E-norm distance from \textbf{any given} Stinespring operator  $V_{\Phi}$ of $\Phi$ to the set of all Stinespring operators of $\Psi$ with the same environment.
}\smallskip

As noted in the Introduction this result is similar to the infinite-dimensional version of Uhlmann's theorem. Indeed, since
any purification in $\H_{AE}$ of an operator  $\sigma$ in $\T_+(\H_A)$ has the form $[I_A\otimes U]|\psi\rangle$, where
$|\psi\rangle$ is a given purifications of $\sigma$ and $U$ is a partial isometry such that $[I_B\otimes U^*U]|\psi\rangle=|\psi\rangle$,
equality (\ref{U-th+}) means that \emph{the Bures distance between operators $\rho$ and $\sigma$ in $\T_+(\H_A)$ is equal to the
Hilbert norm distance from any given  purification $|\varphi\rangle$ of $\rho$ to the set of all purifications of $\sigma$  with the same reference system}.
It is essential that we can not consider the above operator $U$ as unitary.
\smallskip

\begin{remark}\label{main-r} The proof of Theorem \ref{main} involves many technical constructions (the auxilarily quantities $\|\cdot\|_{E,n}^H$,  $\beta_{E,n}^H$, etc.), several approximation steps and separate consideration of the cases $\dim\H_E=+\infty$  and $\dim\H_E<+\infty$, while the basic idea of this proof is quite simple: starting from
the minimax relation
\begin{equation}\label{minmax-r}
\min_{U\in\B_1(\H_E)}\max_{\rho\in\C_{H,\!E}}\sqrt{f(\rho,U)}
=\max_{\rho\in\C_{H,\!E}}\min_{U\in\B_1(\H_E)}\sqrt{f(\rho,U)}.
\end{equation}
where $f(\rho,U)=\Tr\Phi(\rho)+\Tr\Psi(\rho)-2\Re\Tr[I_B\otimes U]V_{\Psi}\rho V^*_{\Phi}$, to show that
\begin{equation}\label{minmax-r+}
\inf_{U\in\mathfrak{W}_{\Psi}}\max_{\rho\in\C_{H,E}}\sqrt{f(\rho,U)}
=\max_{\rho\in\C_{H,\!E}}\min_{U\in\B_1(\H_E)}\sqrt{f(\rho,U)}.
\end{equation}
The similar approach was used in the proof of Theorem 1 in \cite{Kr&W+}. For the first view, to prove this implication it suffices to show that
\begin{equation*}%\label{minmax-r++}
\min_{U\in\mathfrak{W}_{\Psi}}\sqrt{f(\rho,U)}=\min_{U\in\B_1(\H_E)}\sqrt{f(\rho,U)}
\end{equation*}
for any $\rho\in \C_{H,\!E}$, which can be done directly without any auxiliary constructions by the same arguments
as in \cite{Kr&W+}. \emph{But
the subtle point is that it's not sufficient} -- in the Appendix we present an example confirming this conclusion. %Unfortunately, the authors of \cite{Kr&W+} did not comment this point.
In fact, to prove the implication (\ref{minmax-r})$\Rightarrow$(\ref{minmax-r+}) directly  we have to show that for any
saddle point $(\rho_0,U_0)$ of the function $f$ (i.e. such a point that relation (\ref{s-p}) with $f_n=f$ holds) the operator $U_0$
belongs to the set  $\mathfrak{W}_{\Psi}$, but this is not true in general (even in the finite-dimensional case). It was the need to overcome
this problem that led to the technical complication of the proof.
\end{remark}\smallskip

\begin{remark}\label{o-q} The main result of this article (Theorem 1 in Section 3) is proved under the assumption that the
Hamiltonian (energy observable) $H$ has discrete spectrum of finite multiplicity. This assumption is fulfilled, in particular, in the practically important
case when the input quantum system is a multimode quantum oscillator. Naturally, the question arises about the validity of this result for the operator $H$ of the general form. In the presented proof, the above assumption on the spectrum of $H$ is used substantially, it ensures the compactness of the set of states with bounded energy, which guarantees the existence of a saddle point in the minimax problem under consideration. An interesting open question is to find a proof that did not use this assumption.
\end{remark}

\section*{Appendix: a note on the minimax problem}

Let $X$ and $Y$ be compact convex nonempty subsets of linear topological spaces and $f(x,y)$ a continuous function
on $X\times Y$ affine on $x$ and on $y$. Then Ky Fan's minimax theorem \cite[Theorem 3.1]{Simons} implies that
$$
\inf_{y\in Y}\sup_{x\in X} f(x,y)= \sup_{x\in X}\inf_{y\in Y} f(x,y).
$$

\textbf{Proposition A-1.} \emph{The condition $\,\inf_{y\in Y_0} f(x,y)=\inf_{y\in Y} f(x,y)\,$ for each $x\in X$, where $Y_0$ is a proper subset
of $Y$, does not imply the equality
}$$
\inf_{y\in Y_0}\sup_{x\in X} f(x,y)=\inf_{y\in Y}\sup_{x\in X} f(x,y).
$$

\emph{Proof.} Let $X=Y=[-1,+1]$, $Y_0=\{-1,+1\}$ and $f(x,y)=xy$. Then
$$
\sup_{x\in X}\inf_{y\in Y} xy=\sup_{x\in X} (-|x|)=0,\quad
\inf_{y\in Y} \sup_{x\in X} xy=\inf_{y\in Y} |y|=0
$$
and $\,\inf_{y\in Y} xy=-|x|=\inf_{y\in Y_0} xy\,$ for all $x\in X$,  but
$$
\inf_{y\in Y_0} \sup_{x\in X} xy=\inf_{y\in Y_0} |y|=1\neq 0=\inf_{y\in Y} \sup_{x\in X} xy.
$$

\bigskip
\bigskip

I am grateful to A.S.Holevo and S.W.Weis for useful discussion and remarks. Special thanks to the  participants of the workshop "Quantum information, statistics, probability", September, 2018, Steklov Mathematical Institite, Moscow, (especially, to A.Winter) for discussions  that motivated this research.

\end{document}